\documentclass[doublecol]{epl2} 

\usepackage[latin1]{inputenc}
\usepackage{amsmath}
\usepackage{amsfonts}
\usepackage{amssymb}
\usepackage{graphicx}
\usepackage[dvipsnames]{xcolor}
\usepackage[makeroom]{cancel}
\usepackage{pbox}
\usepackage{feynmf}
\usepackage{caption}
\usepackage[export]{adjustbox}

\usepackage{hyperref}

\usepackage{placeins}

\usepackage{bm}
\usepackage[english]{babel}

\usepackage{mathtools}
\allowdisplaybreaks

\title{Driving XXZ spin chains: magnetic-field and boundary effects}


\author{F. Lange\thanks{\email{langef@uni-greifswald.de}} \and S. Ejima\thanks{\email{ejima@physik.uni-greifswald.de}}  \and H. Fehske\thanks{\email{fehske@physik.uni-greifswald.de}} }
\shortauthor{F. Lange \etal}
\institute{Institut f{\"u}r Physik,
             Universit{\"a}t Greifswald,
             D-17487 Greifswald,
             Germany}

\abstract{
Using the time-evolving block decimation, we study the spin transport through spin-1/2 and spin-1 XXZ chains subjected to an external magnetic field and contacted to noninteracting fermionic leads. 
For generic system-lead couplings, the spin conductance exhibits several resonances as a function of the magnetic-field strength.   
In the spin-1/2 but not the spin-1 case, the coupling to the leads can be fine-tuned to reach a conducting fixed point, where the peak structure is washed out and the spin conductance is large throughout the gapless Luttinger-liquid phase. 
For the Haldane phase of the spin-1 chain, we analyse how the spin transport is affected by spin-1/2 edge states, and argue that two-impurity Kondo physics is realised.
}

\begin{document}

\maketitle
The transport properties of interacting nanostructures connected to leads are currently of great interest and hence have been the subject of many numerical and analytical studies~\cite{KaneFisher2,PhysRevB.50.5528,DMRGKubo,PhysRevLett.101.066804,ReviewSchmitteckert,ConductingFixedPoint1}. 
For the most part, the focus has been on the charge transport, which is easier to access experimentally. 
However, there are also proposals to use a spin battery to drive a spin current through a contacted system~\cite{SpinBattery}. 
Here, motivated by the recent development of antiferromagnetic spintronics~\cite{AFMSpintronics}, we numerically investigate such a setup 
for general XXZ spin chains. 
We  use the density-matrix renormalisation group (DMRG)~\cite{White92} in conjunction with the time-evolving block decimation~\cite{PhysRevLett.91.147902} to calculate the spin conductance at zero temperature and small spin bias. 
While isolated XXZ chains have already been studied extensively~\cite{ShastrySutherland,HaldaneChainAsLL,1DTransportIssues,LargeDSpintransport}, we explicitely include the leads in our simulations, which should significantly affect the transport behaviour. There are numerous studies on the spin transport in related models of open spin chains, where the baths are accounted for through driving terms at the outer sites, see Refs.~\cite{Prosen2009} and \cite{PhysRevE.91.030103}, for example. Our model differs in that we use a closed-system description with tight-binding leads. 

We extend our previous work~\cite{PreviousPaper} on the spin-1/2 XXZ chain by also considering the topological spin-1 Haldane chain~\cite{Ha83}, and adding an external magnetic field. 
The spin chain realises different quantum phases, both gapped and gapless, 
depending on the magnitude of the local spins and the strength of the magnetic field. 
One might expect the system to be conducting in the gapless and insulating in the gapped regimes.
As we will demonstrate, however, this only holds in specific cases with fine-tuned parameters. 
In general, the effects of the contacts and the finite size of the spin chain need to be taken into account.

After describing the theoretical model, we discuss separately the spin-1/2 chain with easy-axis anisotropy and the spin-1 chain with isotropic exchange. For the latter, special attention is paid to the Haldane phase, where the spin-1/2 edge states contribute to the transport. 

\section{Setup}
We consider a junction composed of a spin chain and two fermionic leads. 
It is assumed that the system and the leads are 
initially in the ground state of the Hamilton operator $\hat{H}_0 = \hat{H}_{S} + \hat{H}_{L} + \hat{H}_{S-L} + \hat{H}_h$. 
There, the system part
\begin{align}
\hat{H}_{S} &=  J \sum_{j=1}^{N_{S}-1} \left[ \frac{1}{2} \left( \hat{S}_j^+ \hat{S}_{j+1}^- + \hat{S}_j^- \hat{S}_{j+1}^+ \right) + \Delta \hat{S}_j^z \hat{S}_{j+1}^z \right]
\end{align}
is the usual XXZ chain Hamiltonian and the lead part
\begin{align}
 \hat{H}_{L} &= -t  \sum_{a=l,r} \sum_{\sigma = \uparrow, \downarrow} \sum_{j > 0}
 \left[ \hat{c}_{j\sigma a}^{\dagger} \hat{c}_{j+1,\sigma a}^{\phantom{\dagger} } + \hat{c}_{j+1,\sigma a}^{\dagger} \hat{c}_{j\sigma a}^{\phantom{\dagger} } \right] 
\end{align}
describes two half-infinite tight-binding chains of spinful fermions. 
Throughout this work, the ratio between the exchange constant and the hopping parameter is assumed to be $J/t=1$. 
Furthermore, the chemical potential in the leads shall be zero. 
The lead with $a=l$ $(r)$
is exchange-coupled to the first (last) site of the spin chain:
\begin{align}
  \hat{H}_{S-L} &= \frac{J'}{2} \left[ \hat{c}_{1\uparrow l}^{\dagger}\hat{c}_{1 \downarrow l}^{\phantom{\dagger}} \hat{S}_{1}^- + \hat{c}_{1 \downarrow l}^{\dagger}\hat{c}_{1 \uparrow l}^{\phantom{\dagger}} \hat{S}_{1}^+  \right. \nonumber  \\
    & \left.  \hspace*{2cm} + \Delta  (\hat{c}_{1 \uparrow l}^{\dagger}\hat{c}_{1 \uparrow l}^{\phantom{\dagger}} - \hat{c}_{1 \downarrow l}^{\dagger}\hat{c}_{1 \downarrow l}^{\phantom{\dagger}})\hat{S}_{1}^z \right]  \nonumber \\ 
  &  \hspace*{0.4cm} + \frac{J'}{2} \left[ \hat{c}_{1\uparrow r}^{\dagger}\hat{c}_{1 \downarrow r}^{\phantom{\dagger}} \hat{S}_{N_S}^- + \hat{c}_{1 \downarrow r}^{\dagger}\hat{c}_{1 \uparrow r}^{\phantom{\dagger}} \hat{S}_{N_S}^+  \right. \nonumber  \\    & \left.  \hspace*{2cm} + \Delta  (\hat{c}_{1 \uparrow r}^{\dagger}\hat{c}_{1 \uparrow r}^{\phantom{\dagger}} - \hat{c}_{1 \downarrow r}^{\dagger}\hat{c}_{1 \downarrow r}^{\phantom{\dagger}})\hat{S}_{N_S}^z \right] \, .
\end{align}
Lastly, a homogeneous magnetic field in both the system and the leads is taken into account by
\begin{align}
  \hat{H}_h &= -h \Bigg[ \sum_{j = 1}^{N_S} \hat{S}_j^z  + \frac{1}{2}  \sum_{a=l,r} \sum_{j > 0} (\hat{c}_{j\uparrow a}^{\dagger}\hat{c}_{j\uparrow a}^{\phantom{\dagger}} - \hat{c}_{j\downarrow a}^{\dagger}\hat{c}_{j\downarrow a}^{\phantom{\dagger}}) \Bigg] \, .
\end{align}

The Hamiltonian  $\hat{H}_0$ is perturbed by an additional magnetic field that acts only in the left lead so that the time evolution is governed by $\hat{H} = \hat{H}_0 + \hat{H}_V$, where
\begin{align}
\hat{H}_V &= \frac{V}{2}\sum_{j > 0} (\hat{c}_{j\uparrow l}^{\dagger}\hat{c}_{j\uparrow l}^{\phantom{\dagger}} - \hat{c}_{j\downarrow l}^{\dagger}\hat{c}_{j\downarrow l}^{\phantom{\dagger}}) \, .
\end{align}
This term adds  a  spin bias $V$ between the leads and drives a spin current through the system. The spin current $j^z$ in the nonequilibrium steady state defines the spin conductance $G=j^z/V$. For the spin current at the interface with left lead, we use the definition $j^z = \frac{iJ}{2}(\hat{c}_{1\uparrow l}^{\dagger} \hat{c}_{1\downarrow l}^{\phantom{\dagger}} \hat{S}_1^- - \hat{c}_{1\downarrow l}^{\dagger} \hat{c}_{1\uparrow l}^{\phantom{\dagger}} \hat{S}_1^+)$.

To calculate the steady-state spin current that develops after the spin bias is switched on, we follow the  approach described in Ref.~\cite{ReviewSchmitteckert}. 
Instead of the usual matrix-product states, however, we use a tree-tensor network representation where each lead is split into two branches corresponding to the two values of the spin index~\cite{TTNGeneral,TTNImpurity}. 
Since such a tensor network is loopless, we can employ more or less the same techniques as in the matrix-product state approach. 
First, the ground state of $\hat{H}_0$  is calculated with the DMRG. 
This state is then evolved in time according to $\hat{H}$ with the time-evolving block decimation.
In these simulations, the leads have to be truncated to a finite number of sites, and therefore no true steady state can be reached. 
Nevertheless, one can extract an accurate estimate for the spin conductance by extrapolating the behaviour of the spin current to the steady state.  
For the finite leads, we apply 
damped boundary conditions, where the hopping parameter is smoothly set to zero near the boundaries away from the system, which allows for a variable magnetization in the spin chain and the surrounding region~\cite{DMRGKubo}. 
The leads in our simulations have up to 600 sites, and the maximum bond dimension during the time evolution is 500. 

\section{Spin-1/2}
The first case we study is the spin-1/2 XXZ chain with anisotropy parameter $\Delta = 2$. 
In the absence of an external magnetic field, the system is deep in the gapped N\'eel phase where the zero-temperature spin transport is diffusive. 
At some finite magnetic-field strength $h_c$, a quantum phase transition to a gapless Luttinger-liquid (LL) phase takes place and the spin transport becomes ballistic~\cite{ShastrySutherland,Zotos1997}. 
For $\Delta = 2$, we have $h_c/J \simeq 0.40$. 
The LL parameter is $K=1/4$ at $h=h_c$ (following the convention of Ref.~\cite{GiamarchiBook}) and 
increases as $h$ grows further, approaching $K=1$ at the transition to the fully polarised ferromagnetic phase~\cite{GiamarchiBook}. 

The maximum linear spin conductance of the junction is given by the leads as $G_0 = 1/(4\pi)$. 
Because of scattering at the contacts, this ideal spin conductance may only be achieved when the model parameters are fine-tuned to a conducting fixed point, even if the spin chain is in the LL phase~\cite{PreviousPaper}. 
This situation is similar to the transport in fermionic junctions with abrupt change in the parameter, where such conducting fixed points have been studied previously~\cite{ConductingFixedPoint1,ConductingFixedPoint2,ConductingFixedPoint3}. 
Here, we investigate how the contacts affect the magnetic-field dependence of the spin conductance. 

\begin{figure}[t]
\centering
\includegraphics[width=0.9\linewidth]{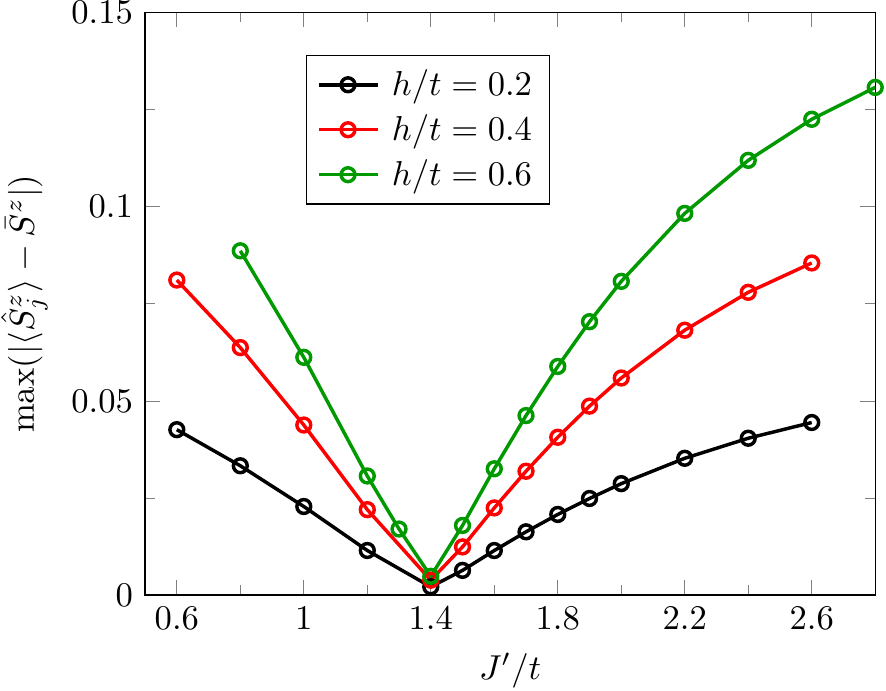}
\caption{Maximum of the Friedel oscillations in the spin-1/2 XXZ chain for $\Delta = 2$ and $\beta t = 4$.}
\label{fig1}
\end{figure}

To confirm that a conducting fixed point exists and to determine the corresponding parameter $J'/t$, we analyse the Friedel oscillations in the spin chain near the interfaces. 
Vanishing Friedel oscillations are expected for perfect contacts, for which the linear spin conductance becomes $G_0$ in the gapless phase~\cite{ConductingFixedPoint1,PreviousPaper}. 
While we are interested in the zero-temperature case, we find it easier to carry out the simulations for finite temperature with the purification method~\cite{MPSPurification}. 
As a simple measure for the strength of the oscillations, the maximum deviation from the average magnetization in the bulk is used. 
The results for inverse temperature $\beta t = 4$ are shown in Fig.~\ref{fig1}. For all field strengths $h/t$ considered, the oscillations approximately vanish at $J'/t=1.4$ which indicates the existence of a conducting fixed point, with no discernible magnetic-field dependence. The temperature dependence of the position of the minimum is negligible for the parameters used. 

\begin{figure}[!t]
\centering
\includegraphics[width=0.9\linewidth]{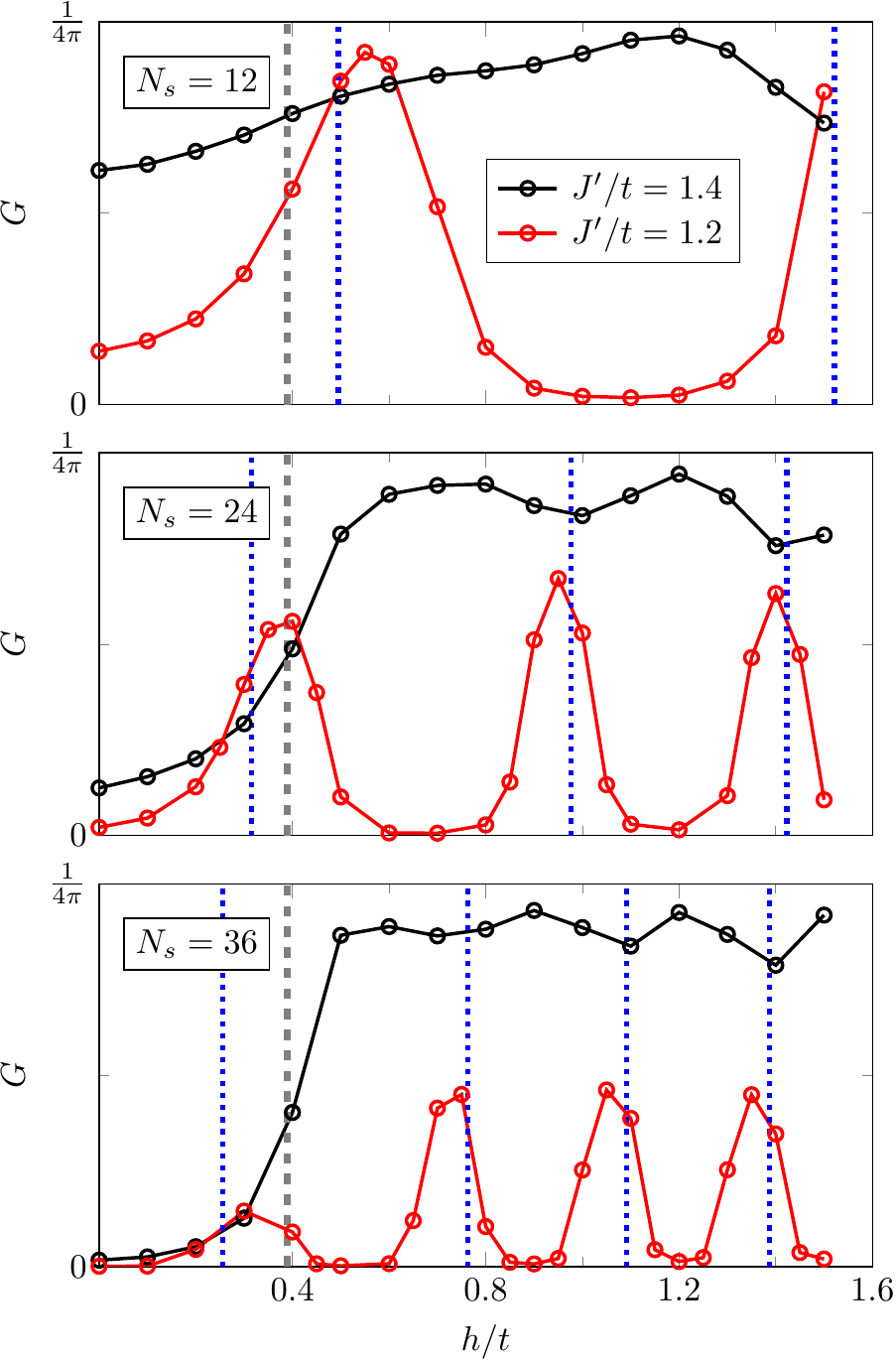}
\caption{Spin conductance $G=j^z/V$ for a spin-1/2-chain junction with $\Delta=2$, and spin bias $V/t=0.1$. The grey lines indicates the critical magnetic field $h_c/t\simeq 0.40$, the blue lines the positions of the level crossings in an isolated spin chain. 
}
\label{fig2}
\end{figure}

We now discuss the effect of the magnetic field and the system-lead coupling on the spin transport. 
In Fig.~\ref{fig2}, the dependence of the spin conductance on the magnetic field strength $h$ is shown for fixed spin bias $V/t=0.1$ and system sizes $N_S = 12, 24$ and $36$. 
Two values of the system-lead coupling are considered: $J'/t=1.4$, where a conducting fixed point is expected, and $J'/t=1.2$. 
For $J'/t=1.4$, conductance stays approximately constant above the critical field $h_c$, except for some fluctuations. The spin conductance there is close to the ideal value $G_0$. 
Below $h_c$, the spin current at fixed $V$ becomes suppressed with increasing chain length $N_S$. 
In the limit of a large spin chain and small bias, the 
spin conductance at the conducting fixed point should be a step function so that the strength of the magnetic field switches between insulating and conducting behaviour: $\lim_{V \rightarrow 0}\lim_{N_S \rightarrow \infty}G = G_0 \Theta(h-h_c)$. Very large spin chains are out of reach of our simulations but the results for $N_S=36$ already resemble this limiting behaviour. 
For a slighly smaller coupling $J'/t=1.2$, away from the conducting fixed point, the effect of the magnetic field 
is qualitatively different. 
Outside of multiple peaks, the spin current is strongly suppressed even for $h>h_c$. 
The number of peaks increases with $N_S$ and their height decreases at fixed $V$. At the first maximum for $N_S=12$, the spin conductance is close to the ideal value $G_0$ while for $N_S=36$ the maximum is approximately $G_0/2$. 
The positions of the current peaks depend strongly on the system size and 
lie roughly at those values of the magnetic field where the ground state of the isolated spin chain becomes degenerate. 
This is in accordance with the picture that even for strong system-lead coupling the spin chain effectively decouples from the leads. 
Similar results were found in Refs.~\cite{DMRGKubo} and \cite{CoulombBlockadeDMRG} for chains of spinless fermions with varying chemical potential. Here, however, the leads are spinful, and states with different magnetization become degenerate. 
The effective low-energy theory near the degeneracy points therefore resembles the two-channel Kondo model, with the two lowest-lying states of the chain corresponding to a spin-1/2~\cite{KondoOddSpinChain}.  
In the zero-bias limit, the two-channel Kondo model exhibits perfect spin conductance~\cite{ChargeKondo}. 
The smaller conductance observed in Fig.~\ref{fig2} may be attributed to additional perturbations in our system and the finite bias $V$. 
In a certain sense, the resonances in the current versus magnetic-field curves correlate with the Coulomb blockade physics known from charge transport through low-dimensional nanostructures~\cite{VanHouten1992}, the $z$-component of the total spin in our model corresponding to the particle number.

\begin{figure}[!t]
\centering
\includegraphics[width=0.9\linewidth]{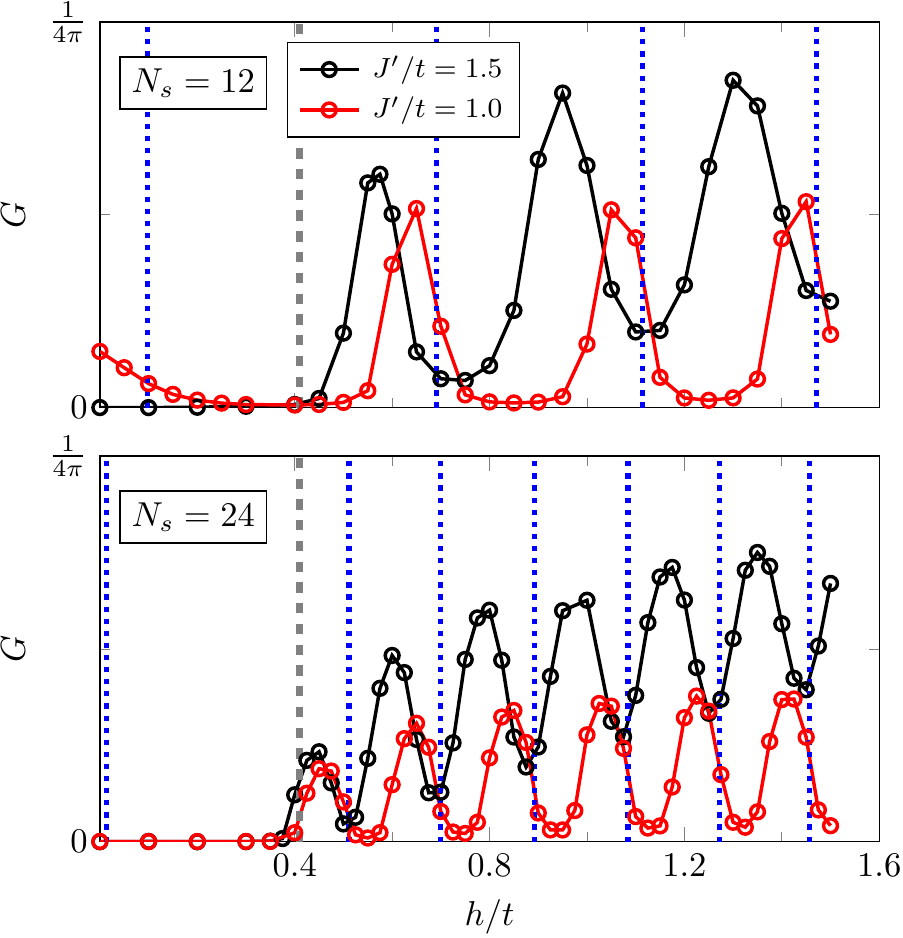}
\caption{Spin conductance for a spin-1-chain junction with $\Delta=1$. The spin bias is $V/t=0.1$. At the critical magnetic field $h_c/t\simeq0.41$ (grey line), the transition between Haldane and Luttinger-liquid phases takes place in the thermodynamic limit. }
\label{fig3}
\end{figure} 

\section{Spin-1}
We now consider a spin-1 chain with $\Delta = 1$, i.e., the Haldane chain. Below the critical magnetic field $h_c/J\simeq 0.41$, the model is in the 
topological Haldane phase with a finite gap for excitations.  
For $h>h_c$, a gapless LL phase is realised. The LL parameter is $K=1$ at the transition, and increases with $h$~\cite{HaldaneChainAsLL}. In contrast to the spin-1/2 chain, the model is nonintegrable so that the linear-response spin transport is expected to be ballistic only at zero temperature~\cite{1DTransportIssues}. 
For small system sizes, the transport behaviour may nevertheless be quite similar to that in an integrable spin-1/2 chain even at finite low temperatures~\cite{LargeDSpintransport}. 
The leads of the junction are the same as in the previous section. 

\begin{figure}[tb]
\centering
\includegraphics[width=0.9\linewidth]{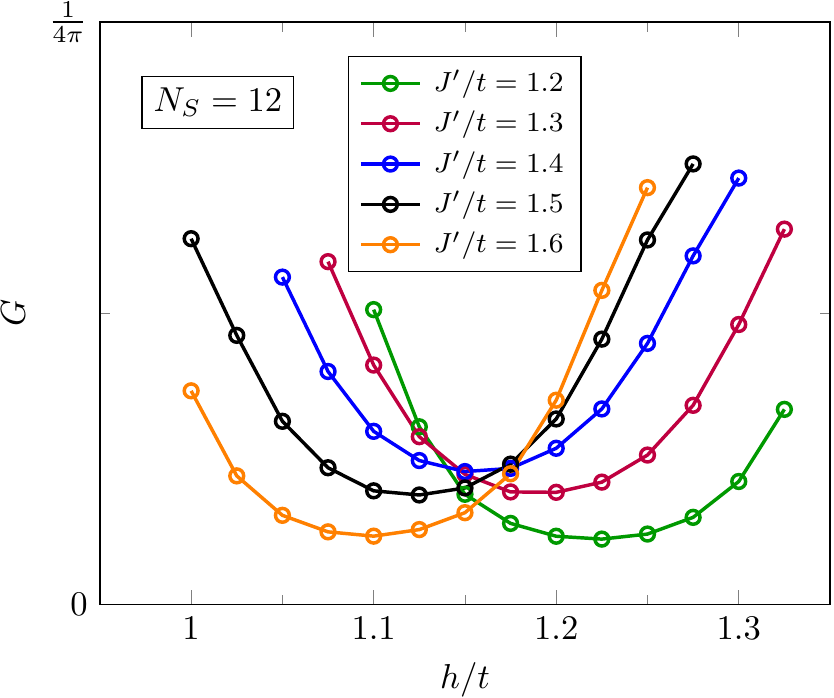}
\caption{Local minimum of the spin conductance for different coupling constants $J'/t$. The other parameters are the same as in Fig.~\ref{fig3}.
}
\label{fig4}
\end{figure}

Let us first discuss the junction for parameters in the LL regime of the spin chain. 
While the bulk of the system for $h>h_c$ can be described by the LL model, the effect of the contacts may be different from the spin-1/2 case. 
Indeed, we find that the Friedel oscillations of the magnetization near the interfaces can not be tuned to zero by adjusting the system-lead coupling $J'$. 
Since the absence of the oscillations indicated a conducting fixed point in the spin-1/2 model, this suggests that here no such fixed point occurs. 
We have checked this by calculating the spin conductance for different values of $J'$ and $h$. The results are displayed in Figs.~\ref{fig3} and \ref{fig4}. 
Like in the spin-1/2 junction away from the conducting fixed point, there are multiple peaks in the spin conductance, which are related to degeneracies in the spectrum of the spin chain. 
As demonstrated in Fig.~\ref{fig4}, however, the minima of the current cannot be removed by tuning the coupling parameter $J'/t$, confirming the absence of a conducting fixed point in the spin-1 junction. 
Although no conducting fixed point is observed, the system-lead coupling still has a significant effect on the conductance as can be seen by comparing the results for $J'/t=1.5$ and $J'/t=1.0$ in Fig.~\ref{fig3}. 
For $J'/t=1.5$, the current values at the minima and maxima become visibly larger with $h$, while the effect is much weaker for $J'/t=1.0$. 
A possible explanation for the growth of the current with $h$ is the increase of the LL parameter $K$ with $h$, which could lead to a less severe effect of the inhomogeneity at the interfaces~\cite{KaneFisher2}. 
Like in the spin-1/2 case, we can compare the position of the conductance peaks with the positions of the level crossings in the isolated spin chain. 
Good agreement is found for the distance between the peaks, but there is a noticeable overall shift to smaller $h$. 

For magnetic fields below $h_c$, in the Haldane regime of the spin chain, the spin conductance vanishes for most of the parameters in Fig.~\ref{fig3}.  
For $N_S=12$ and $J'/t=1.0$, however, there is a clear conductance maximum at $h = 0$. As discussed in the following section, this is likely related to the spin-1/2 edge states in the Haldane phase.

\section{Haldane Phase}

For appropriate combination of system size $N_S$ and system-lead coupling $J'$, a conductance maximum was found around zero magnetic field in the Haldane phase (see Fig.~\ref{fig3}). 
To explain this maximum, one may use an effective low-energy model, where the Hilbert space of the spin chain is reduced to the subspace of the quasidegenerate ground states. 

\begin{table}[!t]
\centering
\begin{tabular}{ll}
$N_S$ & $\tilde{J}/J$ \\
\hline
12 & 0.09714 \\
24 & 0.01341 \\
36 & 0.00186 \\
48 & 0.00025
\end{tabular}
\caption{Strength of the effective exchange coupling between the edge spins calculated from the spin gap for open boundary conditions.}
\label{thetable}
\end{table}

\begin{figure}[!b]
\centering
\includegraphics[width=0.9\linewidth]{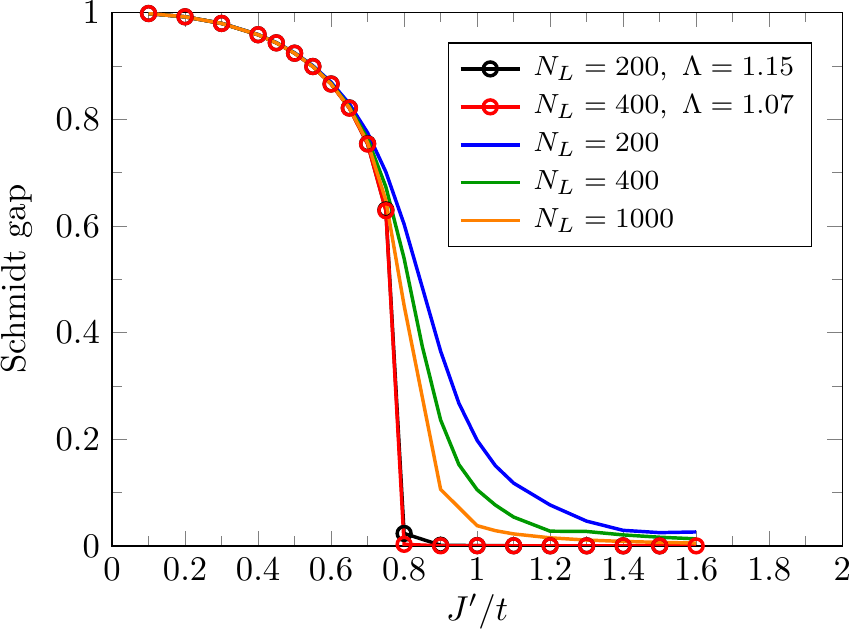}
\caption{Schmidt gap for a junction with $N_S = 12$ and different lead sizes $N_L$ at the bond between the spin chain and the left lead. Results are shown for uniform leads and for logarithmic discretisation.} 
\label{fig5}
\end{figure}

A Haldane chain with open boundary conditions has spin-1/2 degrees of freedom at the edges, which lead to a fourfold-degenerate ground state in the thermodynamic limit. 
When the system size is finite, the edge states are coupled and the degeneracy is partially lifted. 
An effective low-energy model for the spin chain is $\hat{H} = \tilde{J} \hat{\bm{S}}_1 \cdot \hat{\bm{S}}_2$,  
where $\hat{\bm{S}}_{1/2}$ are spin-1/2 operators representing the edge states~\cite{EffectiveEdgeModel1}. The coupling $\tilde{J}$ decreases approximately exponentially with the system size $N_S$ (Table~\ref{thetable}). For even (odd) $N_S$ the coupling is antiferromagnetic (ferromagnetic). Here, we restrict ourselves to even $N_S$. 
Adding the leads results in a spin-1/2 junction with $N_S=2$, exchange coupling $\tilde{J}$ and some effective coupling to leads $\tilde{J}'\propto J'$. By calculating the matrix elements of the spin-1 operators $\hat{\bm{S}}_{1,N_S}$ in the subspace of the quasidegenerate ground states of the isolated spin chain, one obtains the coupling $\tilde{J}' \simeq 1.064 J'$ for $N_S \rightarrow \infty$~\cite{White92,EffectiveEdgeModel2}. 
The resulting spin-1/2 Hamiltonian is a two-impurity two-channel Kondo model~\cite{TwoImpurityKondoAffleck,TwoImpurityAnderson}. For $\tilde{J} > 0$, the model has two phases with an impurity phase transition at $\tilde{J} = \tilde{J}_c$. In the Kondo-singlet phase ($\tilde{J} < \tilde{J}_c$) each spin-1/2 is screened by the corresponding lead and the two spins effectively decouple. In the local-singlet phase ($\tilde{J} > \tilde{J}_c$), the two spins form a singlet and decouple from the leads.

The phase transition $\tilde{J}=\tilde{J}_c$ is called the non-Fermi-liquid fixed point. 
There, the system has some unusual properties such as a residual impurity entropy $\frac{1}{2}\text{ln}(2)$ and a logarithmic divergence in the temperature dependence of the staggered susceptibility~\cite{TwoImpurityKondoAffleck,NRGReview}. 
Since these quantities are difficult to access with the DMRG, we search for a phase transition in the spin-1 model by calculating appropriate entanglement spectra instead~\cite{SchmidtGap}. 
To distinguish between the local-singlet and the Kondo-singlet phase, it is sufficient look at the Schmidt gap which is defined as the difference between the two largest eigenvalues of the reduced density matrix~\cite{SchmidtGap}.

\begin{figure}[!t]
\centering
\includegraphics[width=0.9\linewidth]{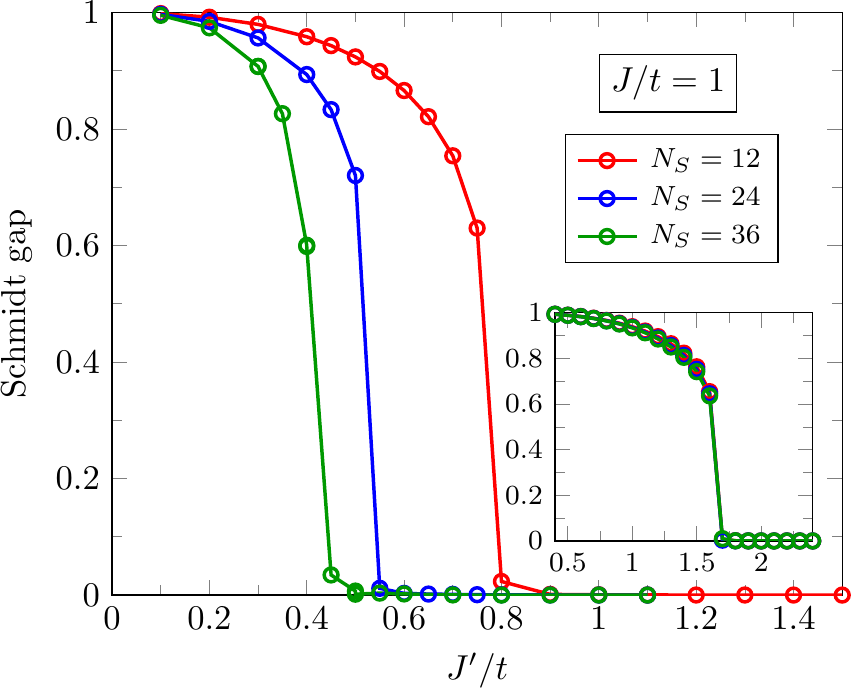}
\caption{Schmidt gap for a spin-1 chain with $\Delta = 1$ at the bond between the spin chain and the left lead. Compared are results for different chain lengths $N_S$. The inset shows the results of a similar calculation for the spin-1/2 chain with $\Delta =1$.} 
\label{fig6}
\end{figure}

\begin{figure}[!t]
\centering
\includegraphics[width=0.9\linewidth]{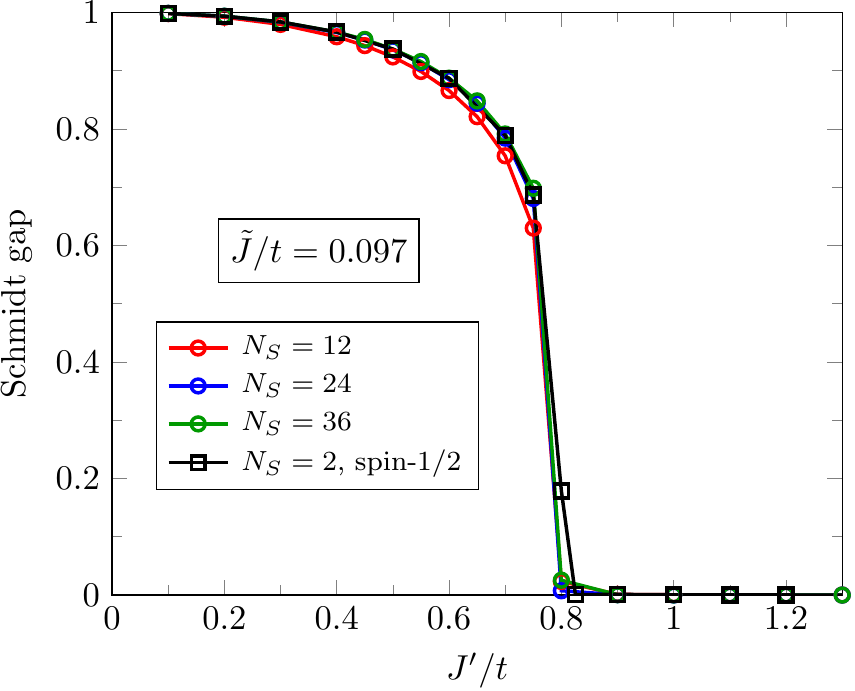}
\caption{Schmidt gap for a spin-1 chain with $\Delta = 1$ for different system sizes and fixed exchange constant for the edge spins $\tilde{J}/t \simeq 0.097$, which corresponds to $N_S=12$ and $J/t=1$ (Table~\ref{thetable}). The results for the assumed effective spin-1/2 chain model are also shown. }
\label{fig7}
\end{figure}

As already exploited in Ref.~\cite{DMRGEnergySpace}, a logarithmic discretization of the leads works well for the calculation of the Schmidt gap (Fig.~\ref{fig5}). The discretization intervals are then defined by the points $\epsilon_n = \pm 2t \Lambda^{-n}$ for $n=0,1,...$ and some discretization parameter $\Lambda > 1$~\cite{NRGReview}. The chain representation of the leads is obtained with the Lanczos algorithm. 
If not mentioned otherwise, leads with $N_L=200$ sites and discretization parameter $\Lambda = 1.15$ are used. 
Figure~\ref{fig6} displays the Schmidt gap for the bond between the spin chain and the left lead in a spin-1 chain. 
There is a sharp transition between finite and zero Schmidt gap whose position changes when the size of the spin chain is varied at fixed exchange coupling $J$. 
If, on the other hand, $J/t$ is adjusted to keep the effective spin-1/2 model constant the position of the jump stays approximately the same (Fig.~\ref{fig7}). The results for the spin-1 chain also agree with a calculation directly in the expected effective spin-1/2 model, which suggests that the observed transition is indeed related to the edge states even though we have no direct confirmation that two-impurity Kondo physics are realised.  
An analogous calculation of the Schmidt gap can be done for a spin-1/2 chain (inset of Fig.~\ref{fig6}), where the jump in the Schmidt gap corresponds to the conducting fixed point. In contrast to the spin-1 case, the position of the jump is approximately independent of the chain length $N_S$ for suffiently large $N_S$.

\begin{figure}[!b]
\centering
\includegraphics[width=0.9\linewidth]{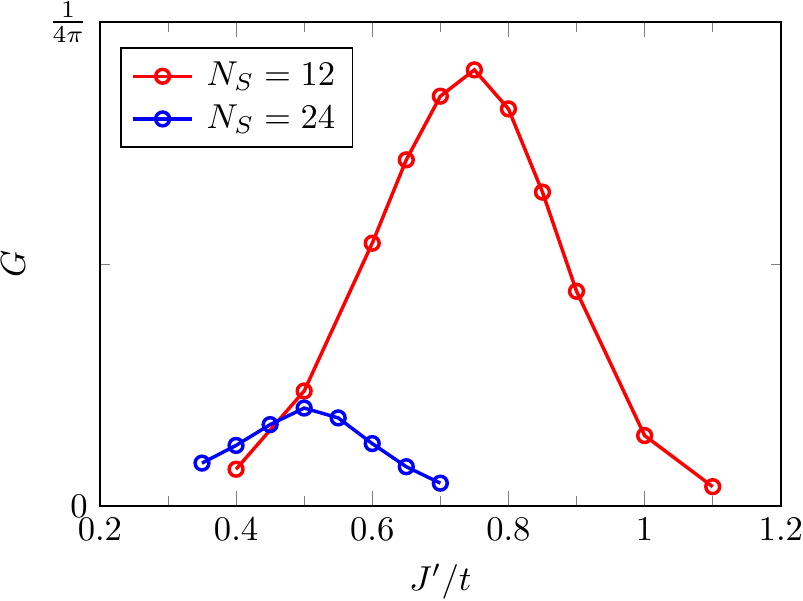}
\caption{Spin conductance for a spin-1 junction with $J/t=1$ and $\Delta=1$. The spin bias is $V/t=0.1$. } 
\label{fig8}
\end{figure}

To see how the spin transport is related to the edge states, the spin conductance is investigated for a spin-1 junction, $h=0$, and different coupling strengths $J'/t$ (Fig.~\ref{fig8}). The spin current is very small except for a peak 
at approximately the position of the jump in the Schmidt gap ($J'/t\simeq 0.75$ for $N_S=12$ and $J'/t\simeq 0.50$ for $N_S=24$). 
With increasing system size $N_S$ the maximum of the conductance at fixed $V$ decreases because of the exponential suppression of the interaction between the edge spins. 
For low spin bias $V$, spin transport in the Haldane chain seems to be possible only near the phase transition between Kondo-singlet and local-singlet phases of the edge states. 
 The spin chain also should be relatively short to have a reasonably strong 
coupling between the edge states. 

When the coupling $J'/t$ is smaller than the critical value and the system is in the local-singlet phase, the maximum of the conductance is shifted to a small finite $h$. This may be explained within an effective single-impurity two-channel Kondo model for a pseudospin formed from the singlet state and the lowest triplet state of the edge spins~\cite{TwoImpurityAnderson}. Accordingly, the conductance maximum lies roughly at $h=J'$, where the singlet and lowest triplet state have the same energy. 
In Fig.~\ref{fig3}, this corresponds to the blue lines below the critical field $h_c$.

\section{Conclusion}
We have numerically studied the spin transport for junctions in which a spin-1/2 or spin-1 XXZ chain is coupled at both ends to noninteracting fermionic leads. 
The focus was on the effect of the contacts and an external magnetic field that switches between gapped and gapless phases of the spin chain. 
When the strength of the magnetic field is varied, the spin conductance generally shows several resonances 
that correspond to degeneracy points in the spectrum of an isolated spin chain. 
The height of these current maxima decreases with the system size, indicating that spin transport becomes suppressed for long chains. 
By fine-tuning the system-lead coupling, a conducting fixed point may be reached. There, instead of showing multiple pronounced peaks, 
the conductance develops a plateau, when the magnetic field is increased and the system passes 
into the gapless phase.  
Whether a conducting fixed point exists will depend on the spin chain in the junction. 
Our results indicate that such a point occurs for the spin-1/2 XXZ chain but not for the spin-1 Haldane chain. A difference is also observed regarding the Friedel oscillations at the contacts, which vanish at the conducting fixed point of the spin-1/2 chain but appear to be always finite in the spin-1 case.   
Lastly, we have discussed the transport in the gapped Haldane phase 
at zero magnetic field. 
The spin conductance there shows a clear maximum as a function of the system-lead coupling. 
By examining the entanglement gap at the interface, we have provided evidence that this conductance maximum corresponds to the non-Fermi-liquid fixed point of an effective two-impurity Kondo model formed by the spin-1/2 edge states.

\section{Acknowledgments}
We thank F. G\"ohmann, T. Shirakawa and S. Yunoki for enlightening discussions. 
DMRG simulations were performed using the ITensor library~\cite{ITensor}. 
F. L. was supported by Deutsche Forschungsgemeinschaft through project FE 398/8-1. 


\begin{thebibliography}{10}
\expandafter\ifx\csname url\endcsname\relax\def\url#1{\texttt{#1}}\fi

\bibitem{KaneFisher2}
\Name{Kane C.~L. \and Fisher M. P.~A.} \REVIEW{Phys. Rev. B}{46}{1992}{15233}.
\newline\url{https://link.aps.org/doi/10.1103/PhysRevB.46.15233}


\bibitem{PhysRevB.50.5528}
\Name{Jauho A.-P., Wingreen N.~S. \and Meir Y.} \REVIEW{Phys. Rev.
  B}{50}{1994}{5528}.
\newline\url{https://link.aps.org/doi/10.1103/PhysRevB.50.5528}


\bibitem{DMRGKubo}
\Name{Bohr D., Schmitteckert P. \and W\"olfle P.} \REVIEW{Europhys.
  Lett.}{73}{2006}{246}.
\newline\url{https://doi.org/10.1209/epl/i2005-10377-6}


\bibitem{PhysRevLett.101.066804}
\Name{Anders F.~B.} \REVIEW{Phys. Rev. Lett.}{101}{2008}{066804}.
\newline\url{https://link.aps.org/doi/10.1103/PhysRevLett.101.066804}


\bibitem{ReviewSchmitteckert}
\Name{Bransch\"adel A., Schneider G. \and Schmitteckert P.} \REVIEW{Ann. Phys.
  (Berlin)}{522}{2010}{657}.
\newline\url{https://onlinelibrary.wiley.com/doi/abs/10.1002/andp.201000017}

\bibitem{ConductingFixedPoint1}
\Name{Sedlmayr N., Ohst J., Affleck I., Sirker J. \and Eggert S.} \REVIEW{Phys.
  Rev. B}{86}{2012}{121302}.
\newline\url{https://link.aps.org/doi/10.1103/PhysRevB.86.121302}

\bibitem{SpinBattery}
\Name{Wang D.-K., Sun Q.-f. \and Guo H.} \REVIEW{Phys. Rev.
  B}{69}{2004}{205312}.
\newline\url{https://link.aps.org/doi/10.1103/PhysRevB.69.205312}

\bibitem{AFMSpintronics}
\Name{Jungwirth T., Marti X., Wadley P. \and Wunderlich J.} \REVIEW{Nat.
  Nanotechnol.}{11}{2016}{231}.
\newline\url{http://dx.doi.org/10.1038/nnano.2016.18}

\bibitem{White92}
\Name{White S.~R.} \REVIEW{Phys. Rev. Lett.}{69}{1992}{2863}.
\newline\url{http://link.aps.org/doi/10.1103/PhysRevLett.69.2863}

\bibitem{PhysRevLett.91.147902}
\Name{Vidal G.} \REVIEW{Phys. Rev. Lett.}{91}{2003}{147902}.
\newline\url{https://link.aps.org/doi/10.1103/PhysRevLett.91.147902}

\bibitem{ShastrySutherland}
\Name{Shastry B.~S. \and Sutherland B.} \REVIEW{Phys. Rev.
  Lett.}{65}{1990}{243}.
\newline\url{https://link.aps.org/doi/10.1103/PhysRevLett.65.243}

\bibitem{HaldaneChainAsLL}
\Name{Konik R.~M. \and Fendley P.} \REVIEW{Phys. Rev. B}{66}{2002}{144416}.
\newline\url{https://link.aps.org/doi/10.1103/PhysRevB.66.144416}

\bibitem{1DTransportIssues}
\Name{Zotos X.} \REVIEW{J. Phys. Soc. Jpn. Suppl.}{74}{2005}{173}.
\newline\url{https://doi.org/10.1143/JPSJS.74S.173}


\bibitem{LargeDSpintransport}
\Name{Psaroudaki C., Herbrych J., Karadamoglou J.,
  Prelov\ifmmode~\check{s}\else \v{s}\fi{}ek P., Zotos X. \and Papanicolaou N.}
  \REVIEW{Phys. Rev. B}{89}{2014}{224418}.
\newline\url{https://link.aps.org/doi/10.1103/PhysRevB.89.224418}



\bibitem{Prosen2009}
\Name{Prosen T. \and \ifmmode \check{Z}\else
  \v{Z}\fi{}nidari\ifmmode~\check{c}\else \v{c}\fi{} M.} \REVIEW{J. Stat.
  Mech.: Theor. Exp.}{2009}{2009}{P02035}.
\newline\url{http://stacks.iop.org/1742-5468/2009/i=02/a=P02035}

\bibitem{PhysRevE.91.030103}
\Name{Lenar\ifmmode \check{c}\else \v{c}\fi{}i\ifmmode~\check{c}\else
  \v{c}\fi{} Z. \and Prosen T.} \REVIEW{Phys. Rev. E}{91}{2015}{030103}.
\newline\url{https://link.aps.org/doi/10.1103/PhysRevE.91.030103}


\bibitem{PreviousPaper}
\Name{Lange F., Ejima S., Shirakawa T., Yunoki S. \and Fehske H.} \REVIEW{Phys.
  Rev. B}{97}{2018}{245124}.
\newline\url{https://link.aps.org/doi/10.1103/PhysRevB.97.245124}

\bibitem{Ha83}
\Name{Haldane F. D.~M.} \REVIEW{Phys. Rev. Lett.}{50}{1983}{1153}.
\newline\url{http://link.aps.org/doi/10.1103/PhysRevLett.50.1153}

\bibitem{TTNGeneral}
\Name{Murg V., Verstraete F., Legeza O. \and Noack R.~M.} \REVIEW{Phys. Rev.
  B}{82}{2010}{205105}.
\newline\url{https://link.aps.org/doi/10.1103/PhysRevB.82.205105}

\bibitem{TTNImpurity}
\Name{Holzner A., Weichselbaum A. \and von Delft J.} \REVIEW{Phys. Rev.
  B}{81}{2010}{125126}.
\newline\url{https://link.aps.org/doi/10.1103/PhysRevB.81.125126}

\bibitem{Zotos1997}
\Name{Zotos X., Naef F. \and Prelovsek P.} \REVIEW{Phys. Rev.
  B}{55}{1997}{11029}.
\newline\url{https://link.aps.org/doi/10.1103/PhysRevB.55.11029}

\bibitem{GiamarchiBook}
\Name{Giamarchi T.} \Book{{Quantum physics in one dimension}} (Clarendon Press,
  Oxford) 2003.
\newline\url{http://dx.doi.org/10.1093/acprof:oso/9780198525004.001.0001}

\bibitem{ConductingFixedPoint2}
\Name{Sedlmayr N., Morath D., Sirker J., Eggert S. \and Affleck I.}
  \REVIEW{Phys. Rev. B}{89}{2014}{045133}.
\newline\url{https://link.aps.org/doi/10.1103/PhysRevB.89.045133}

\bibitem{ConductingFixedPoint3}
\Name{Morath D., Sedlmayr N., Sirker J. \and Eggert S.} \REVIEW{Phys. Rev.
  B}{94}{2016}{115162}.
\newline\url{https://link.aps.org/doi/10.1103/PhysRevB.94.115162}

\bibitem{MPSPurification}
\Name{Verstraete F., Garc\'{\i}a-Ripoll J.~J. \and Cirac J.~I.} \REVIEW{Phys.
  Rev. Lett.}{93}{2004}{207204}.
\newline\url{https://link.aps.org/doi/10.1103/PhysRevLett.93.207204}

\bibitem{CoulombBlockadeDMRG}
\Name{Vasseur G., Weinmann D. \and Jalabert R.~A.} \REVIEW{Eur. Phys. J.
  B}{51}{2006}{267}.
\newline\url{https://doi.org/10.1140/epjb/e2006-00210-2}

\bibitem{KondoOddSpinChain}
\Name{Mitchell A.~K., Logan D.~E. \and Krishnamurthy H.~R.} \REVIEW{Phys. Rev.
  B}{84}{2011}{035119}.
\newline\url{https://link.aps.org/doi/10.1103/PhysRevB.84.035119}

\bibitem{ChargeKondo}
\Name{Mitchell A.~K., Landau L.~A., Fritz L. \and Sela E.} \REVIEW{Phys. Rev.
  Lett.}{116}{2016}{157202}.
\newline\url{https://link.aps.org/doi/10.1103/PhysRevLett.116.157202}


\bibitem{VanHouten1992}
\Name{Van~Houten H., Beenakker C. W.~J. \and Staring A. A.~M.}
  \Book{Coulomb-Blockade Oscillations in Semiconductor Nanostructures}
  (Springer US, Boston, MA) 1992 pp. 167--216.
\newline\url{https://doi.org/10.1007/978-1-4757-2166-9_5}


\bibitem{EffectiveEdgeModel1}
\Name{S\o{}rensen E.~S. \and Affleck I.} \REVIEW{Phys. Rev.
  B}{49}{1994}{15771}.
\newline\url{https://link.aps.org/doi/10.1103/PhysRevB.49.15771}

\bibitem{EffectiveEdgeModel2}
\Name{S\o{}rensen E.~S. \and Affleck I.} \REVIEW{Phys. Rev.
  B}{51}{1995}{16115}.
\newline\url{https://link.aps.org/doi/10.1103/PhysRevB.51.16115}

\bibitem{TwoImpurityKondoAffleck}
\Name{Affleck I., Ludwig A. W.~W. \and Jones B.~A.} \REVIEW{Phys. Rev.
  B}{52}{1995}{9528}.
\newline\url{https://link.aps.org/doi/10.1103/PhysRevB.52.9528}

\bibitem{TwoImpurityAnderson}
\Name{Jayatilaka F.~W., Galpin M.~R. \and Logan D.~E.} \REVIEW{Phys. Rev.
  B}{84}{2011}{115111}.
\newline\url{https://link.aps.org/doi/10.1103/PhysRevB.84.115111}

\bibitem{NRGReview}
\Name{Bulla R., Costi T.~A. \and Pruschke T.} \REVIEW{Rev. Mod.
  Phys.}{80}{2008}{395}.
\newline\url{https://link.aps.org/doi/10.1103/RevModPhys.80.395}

\bibitem{SchmidtGap}
\Name{Bayat A., Johannesson H., Bose S. \and Sodano P.} \REVIEW{Nat.
  Commun.}{5}{2014}{3784}.
\newline\url{http://dx.doi.org/10.1038/ncomms4784}

\bibitem{DMRGEnergySpace}
\Name{Shirakawa T. \and Yunoki S.} \REVIEW{Phys. Rev. B}{93}{2016}{205124}.
\newline\url{https://link.aps.org/doi/10.1103/PhysRevB.93.205124}

\bibitem{ITensor}
{h}ttp://itensor.org/.

\end{thebibliography}
\bibliographystyle{eplbib}

\end{document}